\newcounter{myctr}
\def\myitem{\refstepcounter{myctr}\bibfont\noindent\ifnum\themyctr>9\else\phantom{0}\fi\hangindent17pt\themyctr.\enskip}

\documentclass{ws-ijqi}

\usepackage{braket}

\newtheorem{thm}{Theorem}

\begin{document}

\bibliographystyle{unsrt}

\markboth{Takuya Machida}
{A limit theorem for a splitting distribution of a quantum walk}

\title{A limit theorem for a splitting distribution of a quantum walk}

\author{Takuya Machida}

\address{College of Industrial Technology, Nihon University, Narashino, Chiba 275-8576, Japan\\
machida.takuya@nihon-u.ac.jp}

\maketitle

\begin{abstract}
Discrete-time quantum walks are considered a counterpart of random walks and the study for them has been getting attention since around 2000.
In this paper, we focus on a quantum walk which generates a probability distribution splitting to two parts.
The quantum walker with two coin states spreads at points, represented by integers, and we analyze the chance of finding the walker at each position after it carries out a unitary evolution a lot of times.
The result is reported as a long-time limit distribution from which one can see an approximation to the finding probability.
\end{abstract}


\section{Introduction}
Quantum walks were introduced and have been studied in physics, mathematics, and quantum information theory~\cite{Gudder1988,AharonovDavidovichZagury1993,Meyer1996}.
Aharonov et al.~\cite{AharonovDavidovichZagury1993} presented the notion of quantum walk as a counterpart of random walks and Meyer~\cite{Meyer1996} as a quantum cellular automata.  
Since quantum walks are considered a quantization of random walks which are capable of explaining a lot of stochastic phenomena, large attention is being paid to them.
While they are applied to quantum search algorithms in quantum information theory~\cite{Venegas-Andraca2008,Venegas-Andraca2012}, physicists work on a possibility that quantum walks are applied to topological insulators~\cite{ObuseRyuFurusakiMudry2014}.
In this paper, we see a discrete-time quantum walk whose positions are represented by integer points, and study a probability distribution with which the quantum walker is observed at each position.
The quantum walker distributes in a couple of major parts as it gets updated, and the result for the probability distribution will be supplied in a long-time limit distribution.
The study for the limit distributions of the quantum walks started in 2002~\cite{Konno2002a}, and many types of limit distributions have been discovered~\cite{Venegas-Andraca2012}.
Particularly, the long-time limit distributions play an important role to tell us approximations to the probability distributions when quantum walkers have repeated their evolutions a lot of times.

This paper is organized as follows.
We start with the definition of a quantum walk whose positions are represented by integers in Sec.~\ref{sec:introduction}.
The system of quantum walk is described in a tensor Hilbert space and the walker repeats two kinds of unitary evolutions alternately.
After defining a finding probability to observe the walker at each position, we provide a limit theorem after the walker has got updated a lot of times in Sec.~\ref{sec:limit_theorem}.
The theorem results in a helpful description to know an approximate behavior of the walker. 
The proof is also given in the same section.
In the final section we discuss with past studies and summarize this paper.

\section{Definition of a quantum walk}
\label{sec:introduction}
\begin{figure}[h]
\begin{center}
 \begin{minipage}{60mm}
  \begin{center}
   \includegraphics[scale=0.3]{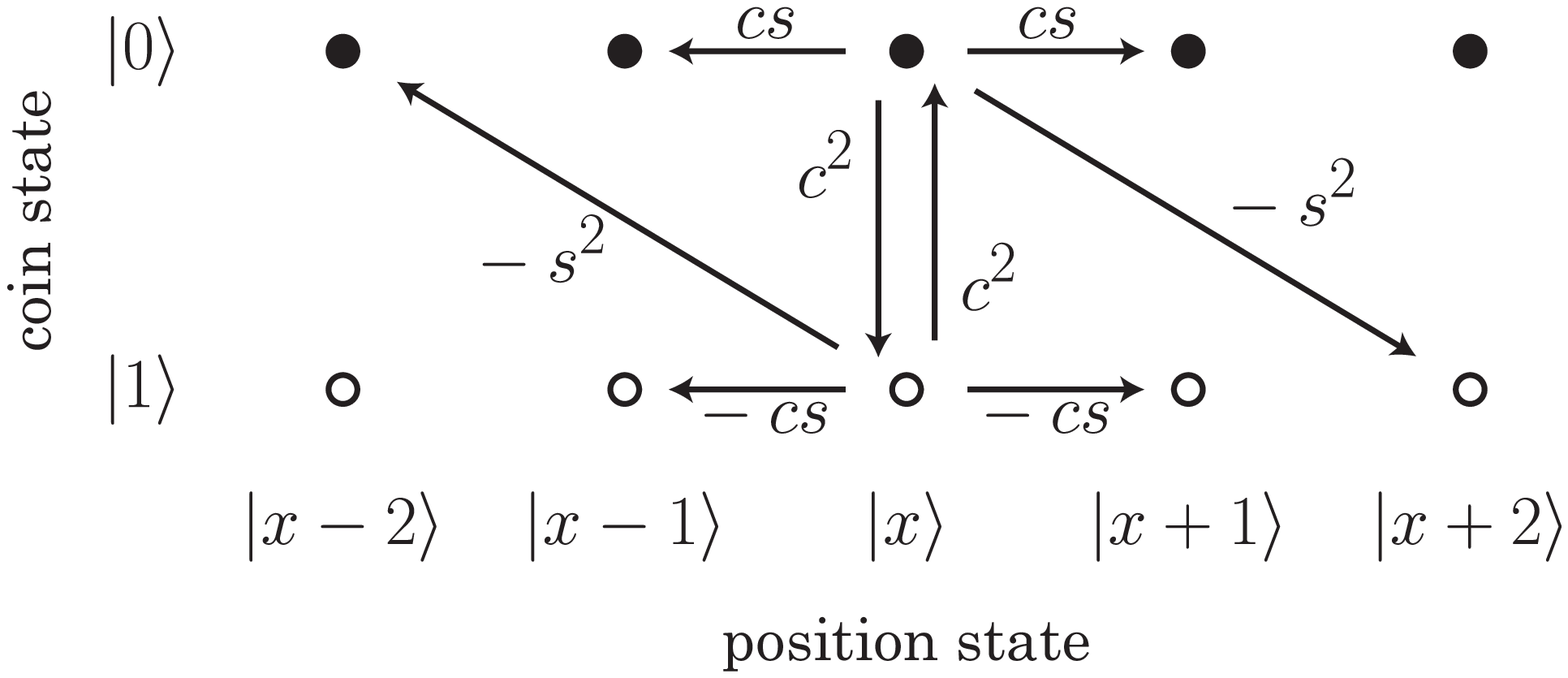}\\[2mm]
  (a) $t=0,2,4,\ldots$
  \end{center}
 \end{minipage}\hspace{5mm}
 \begin{minipage}{60mm}
  \begin{center}
   \includegraphics[scale=0.3]{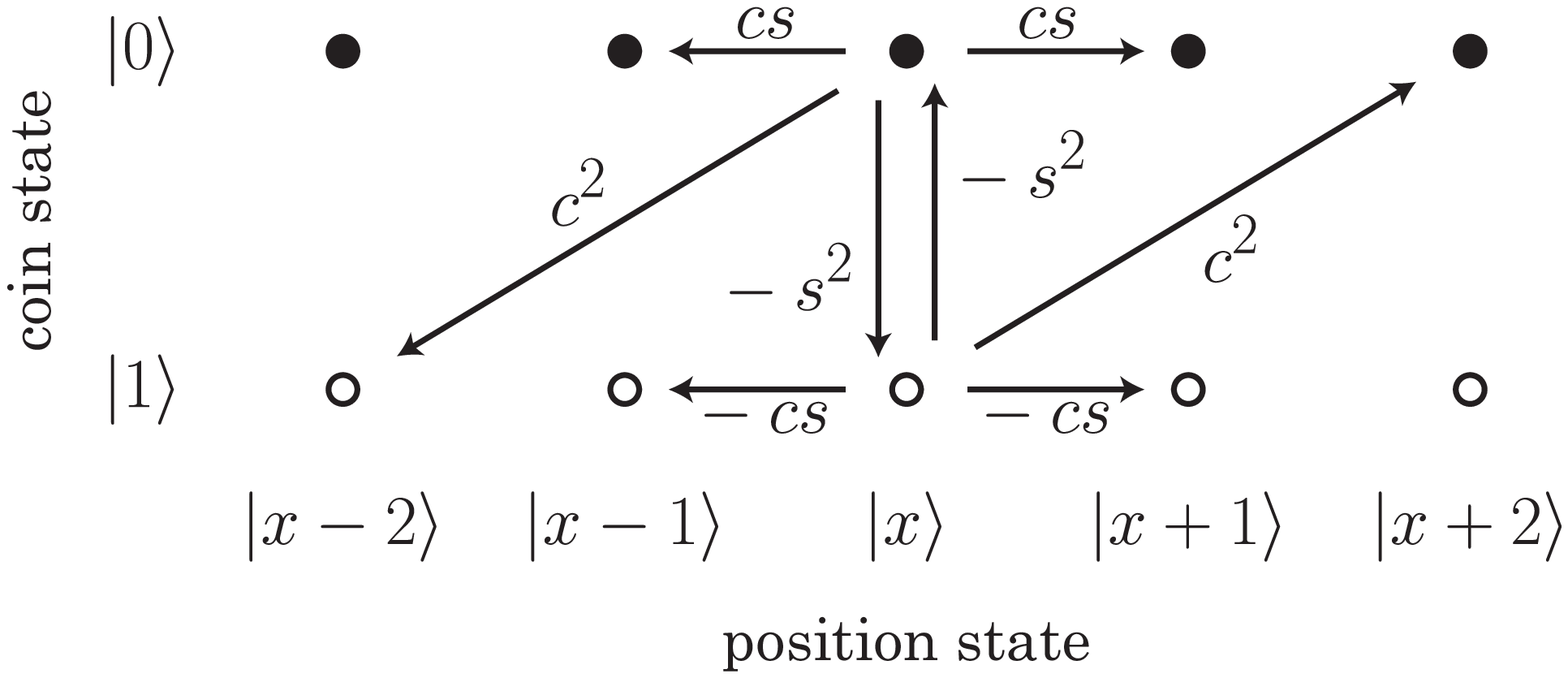}\\[2mm]
  (b) $t=1,3,5,\ldots$
  \end{center}
 \end{minipage}
\caption{$c=\cos\theta, s=\sin\theta$ : The walker at position $x\in\mathbb{Z}$ shifts to other positions and stays at the same position, changing its coin states $\ket{0}$ and $\ket{1}$. The transition is described in Eq.~\eqref{eq:170314_1} with unitary operations $U_1$ and $U_2$.}
\label{fig:3}
\end{center}
\end{figure}

Let us start with the description of a discrete-time quantum walk.
The quantum walker with two coin states $\ket{0}$ and $\ket{1}$ is supposed to locate at points, whose set is represented by $\mathbb{Z}=\left\{0,\pm 1,\pm 2,\ldots\right\}$, in superposition.
Its system is described on a tensor Hilbert space $\mathcal{H}_p\otimes\mathcal{H}_s$.
The Hilbert space $\mathcal{H}_p$ alters the integer points and it is spanned by the orthogonal normalized basis $\left\{\ket{x} : x\in\mathbb{Z}\right\}$.
Also, the Hilbert space $\mathcal{H}_s$ represents the coin states and it is spanned by the orthogonal normalized basis $\left\{\ket{0},\ket{1}\right\}$.
We are, for instance, allowed to define
\begin{equation}
 \ket{0}=\begin{bmatrix}
	  1\\0
	 \end{bmatrix},\quad
	 \ket{1}=\begin{bmatrix}
		  0\\1
		 \end{bmatrix},
\end{equation}
for the Hilbert space $\mathcal{H}_s$.
Quantum walks are defined as unitary processes in which each coin state at each location changes with given unitary operations.
The quantum walker in this paper is also manipulated by unitary operations.
The system of quantum walk at time $t\,(=0,1,2,\ldots)$, represented by $\ket{\Psi_t}\in\mathcal{H}_p\otimes\mathcal{H}_s$, updates with unitary operations $U_1$ and $U_2$ assigned a parameter $\theta\in [0,\pi)$,  
\begin{equation}
 \ket{\Psi_{t+1}}=\left\{\begin{array}{ll}
		   U_1\ket{\Psi_t}& (t=0,2,4,\ldots)\\
			  U_2\ket{\Psi_t}& (t=1,3,5,\ldots)
			 \end{array}\right.,\label{eq:170314_1}
\end{equation}
where
\begin{align}
 U_1=&\sum_{x\in\mathbb{Z}}\ket{x-2}\bra{x}\otimes(-\sin^2\theta)\ket{0}\bra{1}\nonumber\\
 &+\ket{x-1}\bra{x}\otimes\cos\theta\sin\theta\,\Bigl(\ket{0}\bra{0}-\ket{1}\bra{1}\Bigr)\nonumber\\
 &+\ket{x}\bra{x}\otimes\cos^2\theta\,\Bigl(\ket{0}\bra{1}+\ket{1}\bra{0}\Bigr)\nonumber\\
 &+\ket{x+1}\bra{x}\otimes\cos\theta\sin\theta\,\Bigl(\ket{0}\bra{0}-\ket{1}\bra{1}\Bigr)\nonumber\\
 &+\ket{x+2}\bra{x}\otimes (-\sin^2\theta)\ket{1}\bra{0},\\[3mm]
 U_2=&\sum_{x\in\mathbb{Z}}\ket{x-2}\bra{x}\otimes \cos^2\theta\,\ket{1}\bra{0}\nonumber\\
 &+\ket{x-1}\bra{x}\otimes\cos\theta\sin\theta\,\Bigl(\ket{0}\bra{0}-\ket{1}\bra{1}\Bigr)\nonumber\\
 &+\ket{x}\bra{x}\otimes(-\sin^2\theta)\Bigl(\ket{0}\bra{1}+\ket{1}\bra{0}\Bigr)\nonumber\\
 &+\ket{x+1}\bra{x}\otimes\cos\theta\sin\theta\,\Bigl(\ket{0}\bra{0}-\ket{1}\bra{1}\Bigr)\nonumber\\
 &+\ket{x+2}\bra{x}\otimes \cos^2\theta\,\ket{0}\bra{1}.
\end{align}
Figure~\ref{fig:3} visualizes the way the walker at position $x\in\mathbb{Z}$ moves, depending on the value of time $t$.

\begin{figure}[h]
\begin{center}
 \begin{minipage}{60mm}
  \begin{center}
   \includegraphics[scale=0.3]{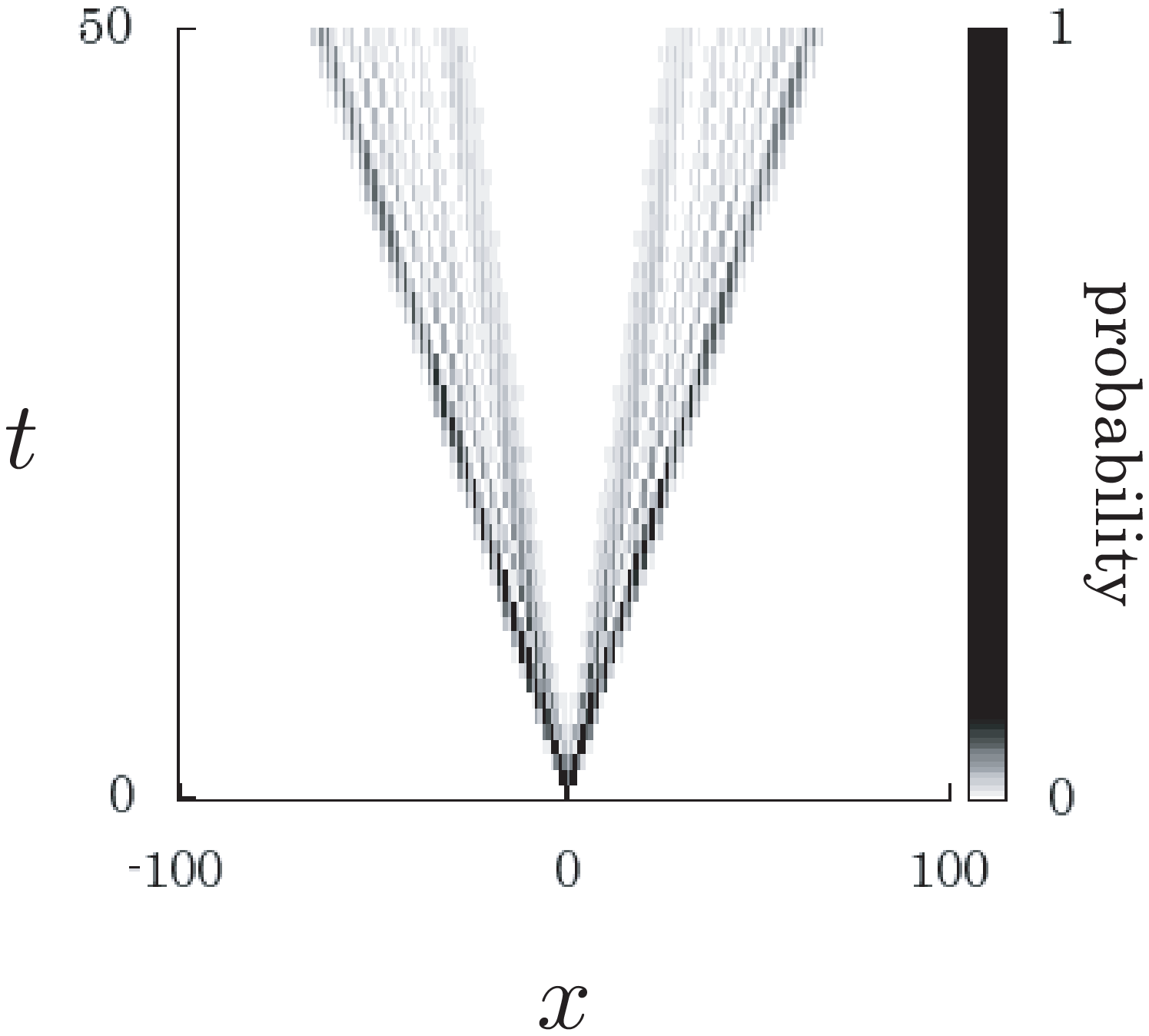}\\[2mm]
  (a) $\theta=\pi/3$
  \end{center}
 \end{minipage}
 \begin{minipage}{60mm}
  \begin{center}
   \includegraphics[scale=0.3]{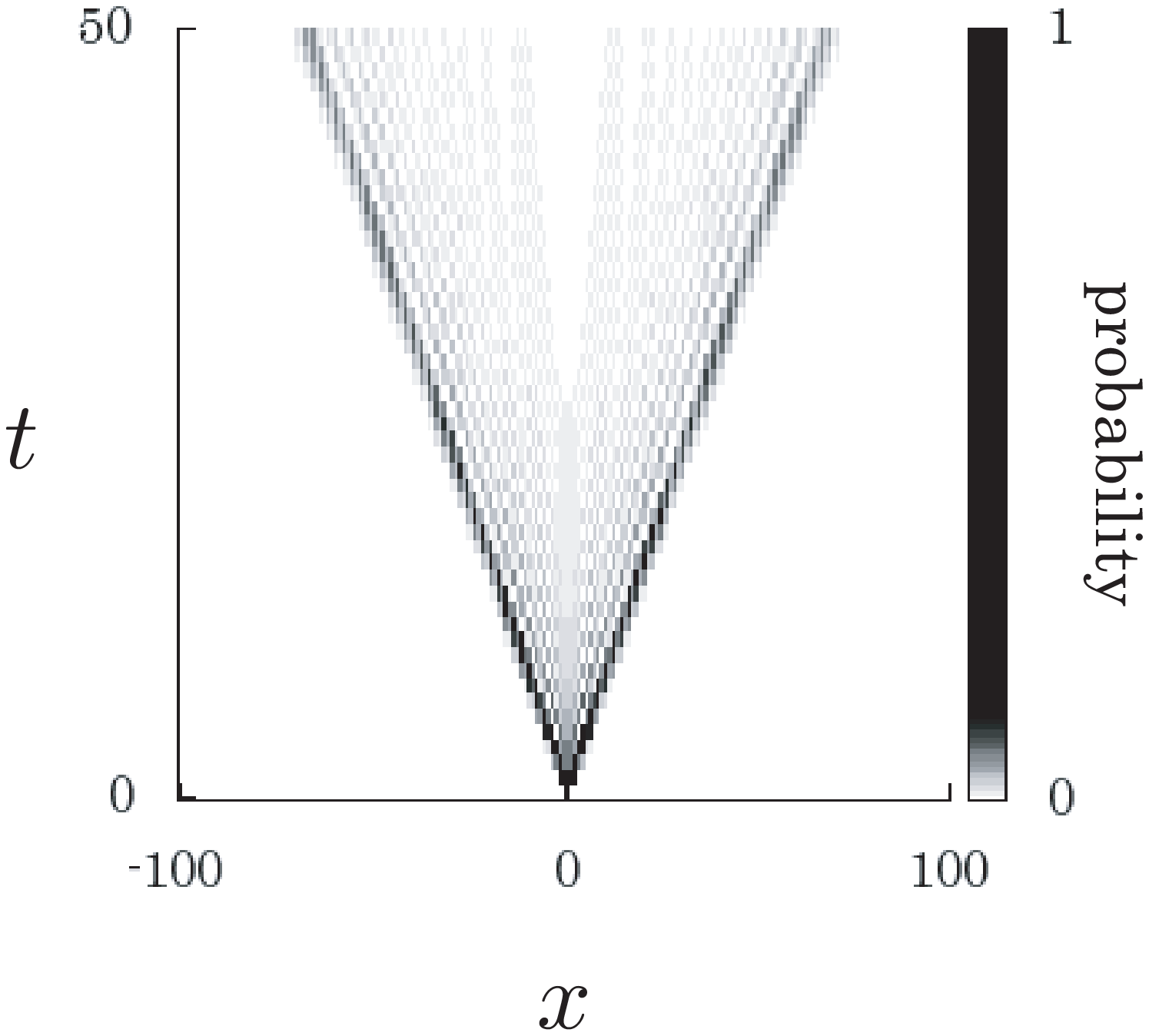}\\[2mm]
  (b) $\theta=\pi/4$
  \end{center}
 \end{minipage}
\caption{The probability distribution $\mathbb{P}(X_t=x)$ can be splitting to two major parts as the walker is getting updated. The walker launches with the localized initial state at the origin, $\ket{\Psi_0}=\ket{0}\otimes (1/\sqrt{2}\ket{0}+i/\sqrt{2}\ket{1})$.}
\label{fig:4}
\end{center}
\end{figure}

\begin{figure}[h]
\begin{center}
 \begin{minipage}{40mm}
  \begin{center}
   \includegraphics[scale=0.2]{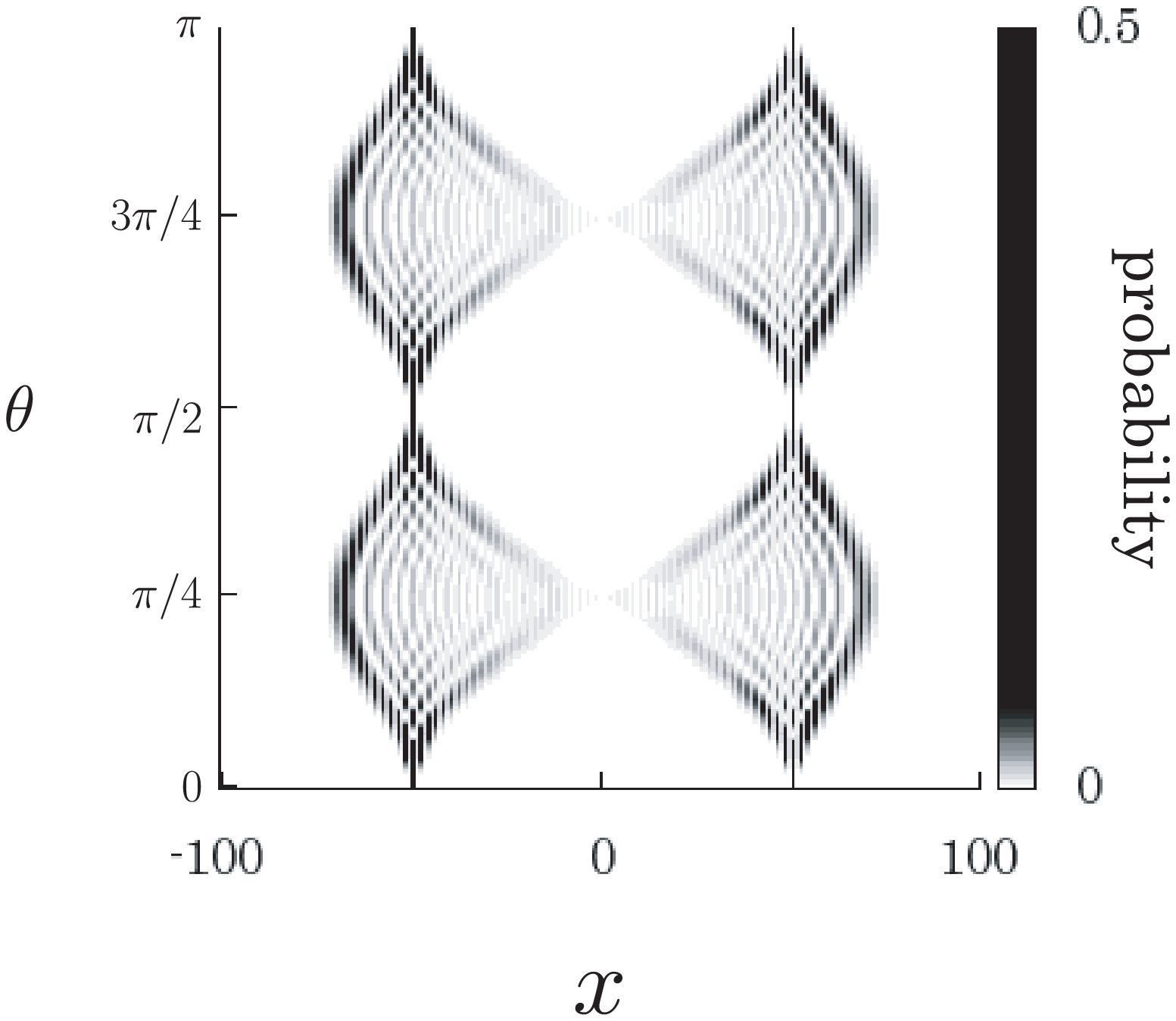}\\[2mm]
  (a) $\alpha=1/\sqrt{2},\, \beta=i/\sqrt{2}$
  \end{center}
 \end{minipage}
 \begin{minipage}{40mm}
  \begin{center}
   \includegraphics[scale=0.2]{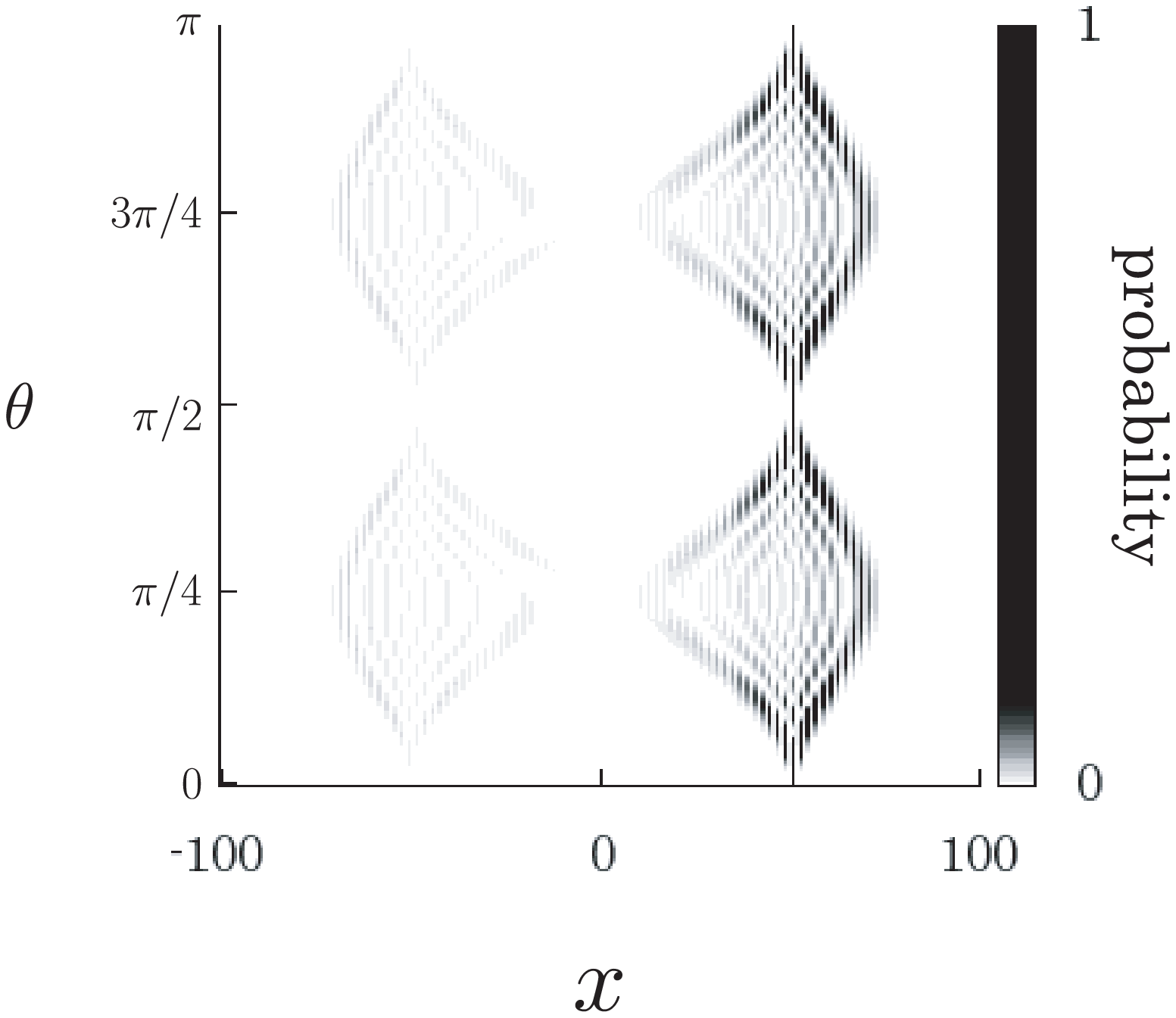}\\[2mm]
  (b) $\alpha=1,\, \beta=0$
  \end{center}
 \end{minipage}
 \begin{minipage}{40mm}
  \begin{center}
   \includegraphics[scale=0.2]{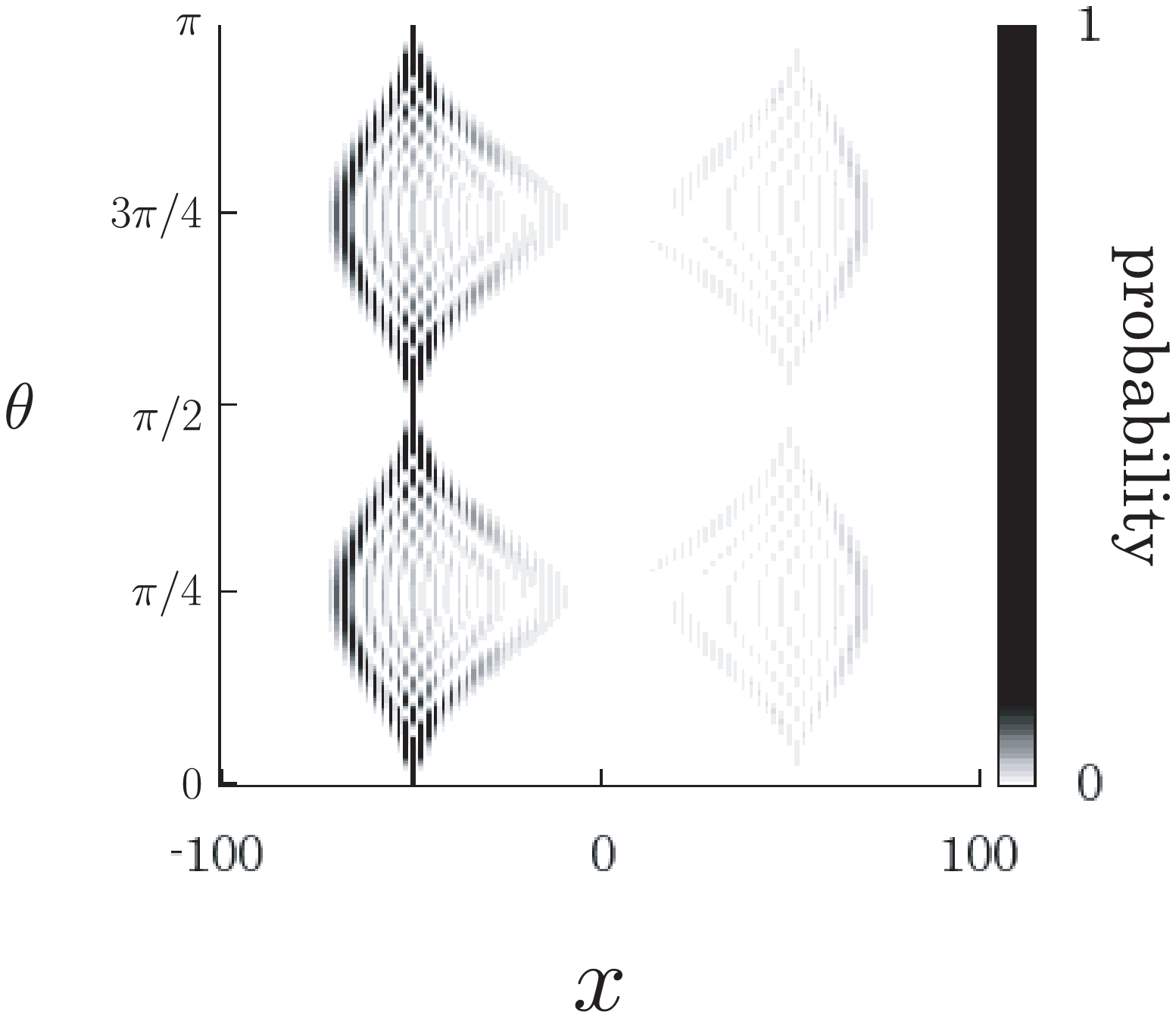}\\[2mm]
  (c) $\alpha=0,\, \beta=1$
  \end{center}
 \end{minipage}
\caption{The probability distribution $\mathbb{P}(X_t=x)$ at time $50$ holds a gap which appears between two major parts, and the width of the gap depends on the value of parameter $\theta$ which is in the unitary operations $U_1$ and $U_2$. The initial state of the walker is given as $\ket{\Psi_0}=\ket{0}\otimes (\alpha\ket{0}+\beta\ket{1})$.}
\label{fig:5}
\end{center}
\end{figure}

We assume in this study that the walker launches with a localized initial state $\ket{\Psi_0}=\ket{0}\otimes\left(\alpha\ket{0}+\beta\ket{1}\right)\,(=\ket{0}\otimes\ket{\phi})$ where the complex numbers $\alpha$ and $\beta$ are supposed to satisfy the constraint $|\alpha|^2+|\beta|^2=1$.
The quantum walker is observed at position $x\in\mathbb{Z}$ at time $t\in\left\{0,1,2,\ldots\right\}$ with probability
\begin{equation}
 \mathbb{P}(X_t=x)=\bra{\Psi_t}\left\{\ket{x}\bra{x}\otimes(\ket{0}\bra{0}+\ket{1}\bra{1})\right\}\ket{\Psi_t},\label{eq:finding_prob}
\end{equation}
where $X_t$ denotes the position of the walker at time $t$.
Figure~\ref{fig:4} depicts how the probability distribution changes as the time $t$ goes up, in which the walker sets off the localized initial state $\ket{0}\otimes(1/\sqrt{2}\ket{0}+i/\sqrt{2}\ket{1})$.
We find the distribution splitting to two major parts in Fig.~\ref{fig:4}-(a).
The width of the gap between the split parts depends on the value of parameter $\theta$ which determines the unitary operations $U_1$ and $U_2$, as shown in Figure~\ref{fig:5}.
We will be able to estimate the width from our main result (Theorem~\ref{th:limit}), visualizing it in Fig.~\ref{fig:2}.

Here, we see the Fourier transform of the quantum walk, which will be used to compute a limit distribution as $t\to\infty$.
Let $i$ be the imaginary unit.
Putting
\begin{align}
 \hat U_1(k)=& 2\cos\theta\sin\theta\cos k\, (\ket{0}\bra{0}-\ket{1}\bra{1})\nonumber\\
 & +(\cos^2\theta-e^{2ik}\sin^2\theta)\,\ket{0}\bra{1}\nonumber\\
 & +(\cos^2\theta-e^{-2ik}\sin^2\theta)\,\ket{1}\bra{0},
\end{align}
\begin{align}
 \hat U_2(k)=& 2\cos\theta\sin\theta\cos k\, (\ket{0}\bra{0}-\ket{1}\bra{1})\nonumber\\
 & +(-\sin^2\theta+e^{-2ik}\cos^2\theta)\,\ket{0}\bra{1}\nonumber\\
 & +(-\sin^2\theta+e^{2ik}\cos^2\theta)\,\ket{1}\bra{0},
\end{align}
we get the evolution of the Fourier transform $\ket{\hat\psi_t(k)}=\sum_{x\in\mathbb{Z}}e^{-ikx}\left\{\bra{x}\otimes(\ket{0}\bra{0}+\ket{1}\bra{1})\right\}\ket{\Psi_t}\, (k\in [-\pi,\pi))$,
\begin{equation}
 \ket{\hat\psi_{t+1}(k)}=\left\{\begin{array}{ll}
			  \hat U_1(k)\ket{\hat\psi_t(k)}& (t=0,2,4,\ldots)\\
				 \hat U_2(k)\ket{\hat\psi_t(k)}& (t=1,3,5,\ldots)
				\end{array}\right.,\label{eq:170314_2}
\end{equation}
from which
\begin{align}
 \ket{\hat\psi_{2t}(k)}=&\left(\hat{U}_2(k)\hat{U}_1(k)\right)^t\ket{\hat\psi_0(k)},\label{eq:F_time_2t}\\
 \ket{\hat\psi_{2t+1}(k)}=&\hat{U}_1(k)\left(\hat{U}_2(k)\hat{U}_1(k)\right)^t\ket{\hat\psi_0(k)},\label{eq:F_time_2t+1}
\end{align}
follow for $t=0,1,2,\ldots$.
Equation~\eqref{eq:170314_2} has come up from Eq.~\eqref{eq:170314_1}.
The initial state of the Fourier transform is computed to be $\ket{\hat\psi_0(k)}=\alpha\ket{0}+\beta\ket{1}$.
We should note that the system is reproduced by inverse Fourier transform
\begin{equation}
 \ket{\Psi_t}=\sum_{x\in\mathbb{Z}}\ket{x}\otimes\int_{-\pi}^\pi\,e^{ikx}\ket{\hat\psi_t(k)}\,\frac{dk}{2\pi}.
\end{equation}

\section{Limit theorem}
\label{sec:limit_theorem}
Analyzing the quantum walk in the Fourier transform, we assert a theorem for the finding probability defined in Eq.~\eqref{eq:finding_prob}.
\begin{thm}
Assume that $\theta\neq 0, \pi/2$.
Let $c$ and $s$ be the short notations for $\cos\theta$ and $\sin\theta$ respectively.
For a real number $x$, we have
 \begin{align}
   &\lim_{t\to\infty}\mathbb{P}\left(\frac{X_t}{t}\leq x\right)\nonumber\\
   =&\int_{-\infty}^x \Bigl\{f(y)\nu_{+}(\alpha,\beta; y)I_\mathcal{D}(y)+f(-y)\nu_{-}(\alpha,\beta; y)I_{\mathcal{D}}(-y)\Bigr\}\,dy,\label{eq:limit_distribution}
 \end{align}
 where
 \begin{align}
  f(x)=&\frac{\left(x+2\sqrt{D(x)}\right)^2}{2\pi(4-x^2)\sqrt{D(x)}\sqrt{W_{+}(x)}\sqrt{W_{-}(x)}},\\[3mm]
  D(x)=&1-16c^4s^4+4c^4s^4x^2,\label{eq:D(x)}\\
  W_{+}(x)=&2(1+4c^2s^2)-(1+2c^2s^2)x^2-x\sqrt{D(x)},\label{eq:W_+}\\
  W_{-}(x)=&-2(1-4c^2s^2)+(1-2c^2s^2)x^2+x\sqrt{D(x)},\label{eq:W_-}
 \end{align}
 \begin{align}
  \nu_{+}(\alpha,\beta; x)=&1+\frac{2c^2s^2x+\sqrt{D(x)}}{1+4c^2s^2}\left(2|\alpha|^2-1\right),\label{eq:nu+}\\
  \nu_{-}(\alpha,\beta; x)=&1+\frac{2c^2s^2x-\sqrt{D(x)}}{1+4c^2s^2}\left(2|\alpha|^2-1\right),\label{eq:nu-}\\
  \mathcal{D}=&\left(\sqrt{1-4c^2s^2},\,\sqrt{1+4c^2s^2}\right),\\
  I_{\mathcal{D}}(x)=&\left\{\begin{array}{cl}
	   1&(x\in \mathcal{D})\\
		  0&(x\notin \mathcal{D})
		 \end{array}\right..
 \end{align}
 \label{th:limit}
\end{thm}

\begin{figure}[h]
\begin{center}
 \begin{minipage}{50mm}
  \begin{center}
   \includegraphics[scale=0.4]{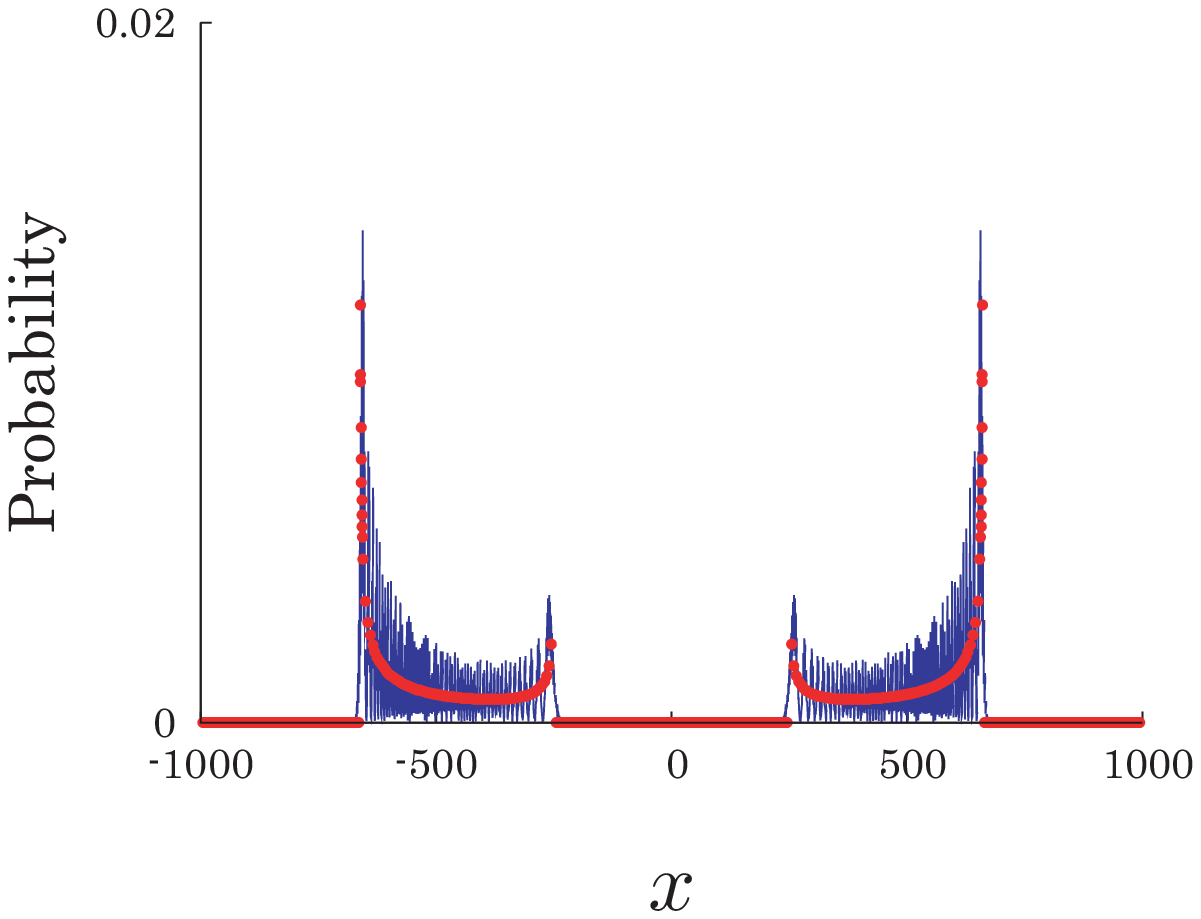}\\[2mm]
  (a) $\theta=\pi/3$
  \end{center}
 \end{minipage}
 \begin{minipage}{50mm}
  \begin{center}
   \includegraphics[scale=0.4]{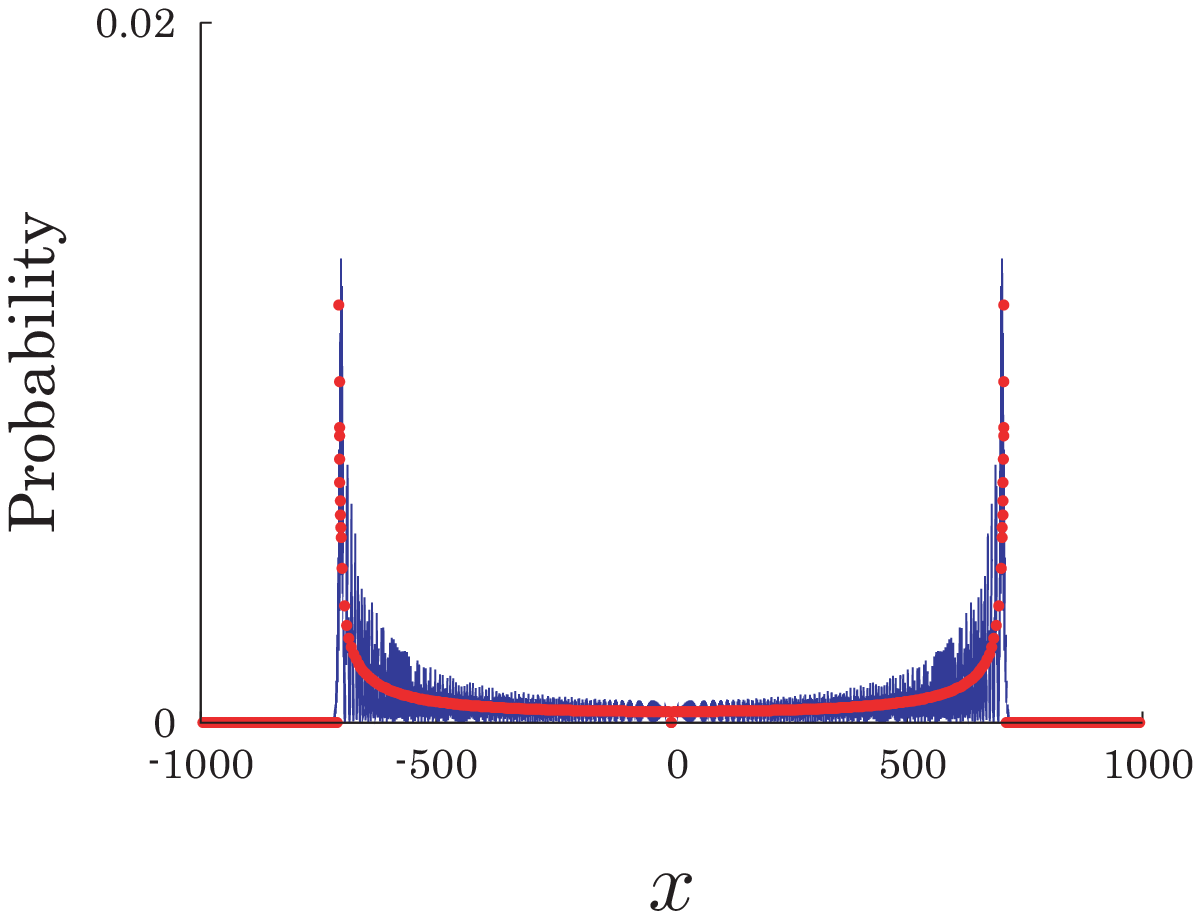}\\[2mm]
  (b) $\theta=\pi/4$
  \end{center}
 \end{minipage}
\caption{The blue lines represent the probability distribution $\mathbb{P}(X_t=x)$ at time $t=500$ and the red points represent the right side of Eq.~\eqref{eq:approximation} as $t=500$. The limit density function approximately reproduces the probability distribution as time $t$ becomes large enough. The walker launches with the localized initial state at the origin, $\ket{\Psi_0}=\ket{0}\otimes (1/\sqrt{2}\ket{0}+i/\sqrt{2}\ket{1})$.}
\label{fig:2}
\end{center}
\end{figure}

The limit distribution is obtained from the convergence of the $r$-th moments $\mathbb{E}[(X_t/t)^r]\,(r=0,1,2,\ldots)$ as $t\to\infty$, and the convergence can be computed by Fourier analysis.
The method for the computation of the long-time limit distributions by Fourier analysis was used to quantum walks in 2004 for the first time~\cite{GrimmettJansonScudo2004} and it has been useful to find limit theorems~\cite{Machida2016a}.
The $r$-th moments of $X_t$ have a representation with the Fourier transform $\ket{\hat\psi_t(k)}$,
\begin{equation}
 \mathbb{E}[X_t^r]=\int_{-\pi}^\pi\,\bra{\hat\psi_t(k)}\left(i^r\frac{d^r}{dk^r}\ket{\hat\psi_t(k)}\right)\,\frac{dk}{2\pi}.
\end{equation}
Recalling Eqs.~\eqref{eq:F_time_2t} and \eqref{eq:F_time_2t+1}, we express the Fourier transform $\ket{\hat\psi_t(k)}$ on the eigenspace of the unitary operation $\hat{U}_2(k)\hat{U}_1(k)$. 
The operation $\hat U_2(k) \hat U_1(k)$ has two eigenvalues, represented by $\lambda_j(k)\,(j=1,2)$, and they are of the form $\lambda_j(k)=g(k)-(-1)^j\,i\sqrt{1-g(k)^2}$ with $g(k)=2c^2s^2\sin^2 2k+\cos 2k$.
We, moreover, hold one of the expressions for the normalized eigenvectors $\ket{v_j(k)}\,(j=1,2)$ associated to the eigenvalues $\lambda_j(k)$,
\begin{align}
 \ket{v_j(k)}=& \frac{1}{\sqrt{N_j(k)}}\biggl[i\Bigl\{(c^4+s^4-2c^2s^2\cos 2k)\sin 2k +(-1)^j\sqrt{1-g(k)^2}\Bigr\}\biggr]\ket{0}\nonumber\\
 & +\frac{1}{\sqrt{N_j(k)}}\biggl[2cs\cos k\Bigl\{1-\cos 2k -i(c^2-s^2)\sin 2k\Bigr\}\biggr]\ket{1},
\end{align}
where the normalized factors are computed to be
\begin{align}
 N_j(k)=&\Bigl\{(c^4+s^4-2c^2s^2\cos 2k)\sin 2k +(-1)^j\sqrt{1-g(k)^2}\Bigr\}^2\nonumber\\
 &+4c^2s^2\cos^2 k \Bigl\{(1-\cos 2k)^2 +(c^2-s^2)^2\sin^2 2k\Bigr\}.
\end{align}
The decomposition of the initial state $\ket{\hat\psi_0(k)}=\sum_{j=1}^2 \braket{v_j(k)|\phi}\ket{v_j(k)}$ gives the representations
\begin{align}
 \ket{\hat\psi_{2t}(k)}=& \sum_{j=1}^2 \lambda_j(k)^t \braket{v_j(k)|\phi}\ket{v_j(k)},\\
 \ket{\hat\psi_{2t+1}(k)}=& \hat{U}_1(k)\sum_{j=1}^2 \lambda_j(k)^t \braket{v_j(k)|\phi}\ket{v_j(k)},
\end{align}
from which
\begin{align}
 \frac{d^r}{dk^r}\ket{\hat\psi_{2t}(k)}=& \biggl\{(t)_r \sum_{j=1}^2 \lambda_j(k)^{t-r} \left(\lambda'_j(k)\right)^r\braket{v_j(k)|\phi}\ket{v_j(k)}\biggr\} + O(t^{r-1}),\\
 \frac{d^r}{dk^r}\ket{\hat\psi_{2t+1}(k)}=& \hat{U}_1(k) \biggl\{(t)_r \sum_{j=1}^2 \lambda_j(k)^{t-r} \left(\lambda'_j(k)\right)^r\braket{v_j(k)|\phi}\ket{v_j(k)}\biggr\}\nonumber\\
 & + O(t^{r-1}),
\end{align}
follow with $(t)_r=t(t-1)\times\cdots\times (t-r-1)$.
We finally reach the limits
\begin{align}
 & \lim_{t\to\infty}\mathbb{E}\left[\left(\frac{X_{2t}}{2t}\right)^r\right]=\lim_{t\to\infty}\mathbb{E}\left[\left(\frac{X_{2t+1}}{2t+1}\right)^r\right]\nonumber\\
 =& \int_{-\pi}^\pi \sum_{j=1}^2 \left(\frac{i\lambda'_j(k)}{2\lambda_j(k)}\right)^r \Bigl|\braket{v_j(k)|\phi}\Bigr|^2\,\frac{dk}{2\pi},\label{eq:170703_1}
\end{align}
where the function $i\lambda'_j(k)/2\lambda_j(k)$ is organized to be of the form
\begin{equation}
 \frac{i\lambda'_j(k)}{2\lambda_j(k)}= (-1)^j\,\frac{\sin 2k}{\,|\sin 2k|\,}\frac{1-4c^2s^2\cos 2k}{\sqrt{1-4c^2s^2(c^2s^2\sin^2 2k+\cos 2k)}}\quad (j=1,2).
\end{equation}
Defining the function
\begin{equation}
 h(k)=\frac{1-4c^2s^2\cos 2k}{\sqrt{1-4c^2s^2(c^2s^2\sin^2 2k+\cos 2k)}},
\end{equation}
we have
\begin{align}
 & \int_{-\pi}^\pi \left(\frac{i\lambda_1'(k)}{2\lambda_1(k)}\right)^r\Bigl|\braket{v_1(k)|\phi}\Bigr|^2\,dk\nonumber\\[2mm]
 =& \int_{-\pi}^{-\frac{\pi}{2}}(-h(k))^r\Bigl|\braket{v_1(k)|\phi}\Bigr|^2\,dk
 + \int_{-\frac{\pi}{2}}^{0} h(k)^r\Bigl|\braket{v_1(k)|\phi}\Bigr|^2\,dk\nonumber\\
 &+ \int_{0}^{\frac{\pi}{2}}(-h(k))^r\Bigl|\braket{v_1(k)|\phi}\Bigr|^2\,dk
 + \int_{\frac{\pi}{2}}^{\pi} h(k)^r\Bigl|\braket{v_1(k)|\phi}\Bigr|^2\,dk\nonumber\\[2mm]
 =& \int_{\frac{\pi}{2}}^{\pi}(-h(-k))^r\Bigl|\braket{v_1(-k)|\phi}\Bigr|^2\,dk
 + \int_{0}^{\frac{\pi}{2}} h(-k)^r\Bigl|\braket{v_1(-k)|\phi}\Bigr|^2\,dk\nonumber\\
 &+ \int_{0}^{\frac{\pi}{2}}(-h(k))^r\Bigl|\braket{v_1(k)|\phi}\Bigr|^2\,dk
 + \int_{\frac{\pi}{2}}^{\pi} h(k)^r\Bigl|\braket{v_1(k)|\phi}\Bigr|^2\,dk\nonumber\\[2mm]
 =& \int_{\frac{\pi}{2}}^{\pi}(-h(k))^r\Bigl|\braket{\overline{v_2(k)}|\phi}\Bigr|^2\,dk
 + \int_{0}^{\frac{\pi}{2}} h(k)^r\Bigl|\braket{\overline{v_2(k)}|\phi}\Bigr|^2\,dk\nonumber\\
 &+ \int_{0}^{\frac{\pi}{2}}(-h(k))^r\Bigl|\braket{v_1(k)|\phi}\Bigr|^2\,dk
 + \int_{\frac{\pi}{2}}^{\pi} h(k)^r\Bigl|\braket{v_1(k)|\phi}\Bigr|^2\,dk\nonumber\\[2mm]
 =& \int_{0}^{\frac{\pi}{2}}(-h(\pi-k))^r\Bigl|\braket{\overline{v_2(\pi-k)}|\phi}\Bigr|^2\,dk
 + \int_{0}^{\frac{\pi}{2}} h(k)^r\Bigl|\braket{\overline{v_2(k)}|\phi}\Bigr|^2\,dk\nonumber\\
 &+ \int_{0}^{\frac{\pi}{2}}(-h(k))^r\Bigl|\braket{v_1(k)|\phi}\Bigr|^2\,dk
 + \int_{0}^{\frac{\pi}{2}} h(\pi-k)^r\Bigl|\braket{v_1(\pi-k)|\phi}\Bigr|^2\,dk\nonumber\\[2mm]
 =& \int_{0}^{\frac{\pi}{2}}(-h(k))^r\Bigl|\braket{v_1(k)|\tilde{\phi}}\Bigr|^2\,dk
 + \int_{0}^{\frac{\pi}{2}} h(k)^r\Bigl|\braket{\overline{v_2(k)}|\phi}\Bigr|^2\,dk\nonumber\\
 &+ \int_{0}^{\frac{\pi}{2}}(-h(k))^r\Bigl|\braket{v_1(k)|\phi}\Bigr|^2\,dk
 + \int_{0}^{\frac{\pi}{2}} h(k)^r\Bigl|\braket{\overline{v_2(k)}|\tilde{\phi}}\Bigr|^2\,dk,
\end{align}
with $\ket{\tilde{\phi}}=\Bigl(\ket{0}\bra{0}-\ket{1}\bra{1}\Bigr)\ket{\phi}$.
The similar computation performs
\begin{align}
 & \int_{-\pi}^\pi \left(\frac{i\lambda_2'(k)}{2\lambda_2(k)}\right)^r\Bigl|\braket{v_2(k)|\phi}\Bigr|^2\,dk\nonumber\\[2mm]
 =& \int_{0}^{\frac{\pi}{2}} h(k)^r\Bigl|\braket{v_2(k)|\tilde{\phi}}\Bigr|^2\,dk
 + \int_{0}^{\frac{\pi}{2}} (-h(k))^r\Bigl|\braket{\overline{v_1(k)}|\phi}\Bigr|^2\,dk\nonumber\\
 &+ \int_{0}^{\frac{\pi}{2}} h(k)^r\Bigl|\braket{v_2(k)|\phi}\Bigr|^2\,dk
 + \int_{0}^{\frac{\pi}{2}} (-h(k))^r\Bigl|\braket{\overline{v_1(k)}|\tilde{\phi}}\Bigr|^2\,dk,
\end{align}
and finally a representation comes out,
\begin{align}
 & \int_{-\pi}^\pi \sum_{j=1}^2 \left(\frac{i\lambda'_j(k)}{2\lambda_j(k)}\right)^r \Bigl|\braket{v_j(k)|\phi}\Bigr|^2\,\frac{dk}{2\pi}\nonumber\\
 =& \int_0^{\frac{\pi}{2}} (-h(k))^r\biggl\{\Bigl|\braket{v_1(k)|\phi}\Bigr|^2 + \Bigl|\braket{\overline{v_1(k)}|\phi}\Bigr|^2 + \Bigl|\braket{v_1(k)|\tilde{\phi}}\Bigr|^2 + \Bigl|\braket{\overline{v_1(k)}|\tilde{\phi}}\Bigr|^2\biggr\}\,\frac{dk}{2\pi}\nonumber\\
 & + \int_0^{\frac{\pi}{2}} h(k)^r\biggl\{\Bigl|\braket{v_2(k)|\phi}\Bigr|^2 + \Bigl|\braket{\overline{v_2(k)}|\phi}\Bigr|^2 + \Bigl|\braket{v_2(k)|\tilde{\phi}}\Bigr|^2 + \Bigl|\braket{\overline{v_2(k)}|\tilde{\phi}}\Bigr|^2\biggr\}\,\frac{dk}{2\pi}.
\end{align}
Putting $h(k)=x$, we achieve a desired representation of the convergence
\begin{align}
 &\lim_{t\to\infty}\mathbb{E}\left[\left(\frac{X_t}{t}\right)^r\right]\nonumber\\
 =&\int_{\sqrt{1-4c^2s^2}}^{\sqrt{1+4c^2s^2}}(-x)^r f(x)\nu_{-}(\alpha,\beta; -x)\,dx \,+\, \int_{\sqrt{1-4c^2s^2}}^{\sqrt{1+4c^2s^2}} x^r f(x)\nu_{+}(\alpha,\beta; x)\,dx\nonumber\\[2mm]
 =&\int_{-\sqrt{1+4c^2s^2}}^{-\sqrt{1-4c^2s^2}} x^r f(-x)\nu_{-}(\alpha,\beta; x)\,dx \,+\, \int_{\sqrt{1-4c^2s^2}}^{\sqrt{1+4c^2s^2}} x^r f(x)\nu_{+}(\alpha,\beta; x)\,dx\nonumber\\[2mm]
 =&\int_{-\infty}^{\infty} x^r f(-x)\nu_{-}(\alpha,\beta; x)I_{\mathcal{D}}(-x)\,dx \,+\, \int_{-\infty}^{\infty} x^r f(x)\nu_{+}(\alpha,\beta; x)I_{\mathcal{D}}(x)\,dx\nonumber\\[2mm]
 =&\int_{-\infty}^\infty x^r \Bigl\{f(-x)\nu_{-}(\alpha,\beta; x)I_{\mathcal{D}}(-x)+f(x)\nu_{+}(\alpha,\beta; x)I_\mathcal{D}(x)\Bigr\}\,dx.
\end{align}
and it guarantees the limit distribution Eq.~\eqref{eq:limit_distribution}.
Note that the equation $h(k)=x\,(k\in (0,\,\pi/2))$ can be solved in the form
\begin{equation}
 k=\frac{1}{2}\arccos\left(\frac{2-x^2-x\sqrt{D(x)}}{2c^2s^2(4-x^2)}\right),
\end{equation}
and its derivative is computed to be
\begin{equation}
 \frac{dk}{dx}=\frac{\left(x+2\sqrt{D(x)}\right)^2}{2(4-x^2)\sqrt{D(x)}\sqrt{W_{+}(x)}\sqrt{W_{-}(x)}}.
\end{equation}
See Eqs.~\eqref{eq:D(x)}, \eqref{eq:W_+}, and \eqref{eq:W_-} about the functions $D(x)$, $W_{+}(x)$, and $W_{-}(x)$.
The limit density function reproduces the finding probability $\mathbb{P}(X_t=x)\,(x\in\mathbb{Z})$ as $t\to\infty$ in approximation,
\begin{align}
 &\mathbb{P}(X_t=x)\nonumber\\
 \sim & \frac{1}{t}\biggl\{f\left(\frac{x}{t}\right)\nu_{+}\left(\alpha,\beta; \frac{x}{t}\right)I_\mathcal{D}\left(\frac{x}{t}\right)+f\left(-\frac{x}{t}\right)\nu_{-}\left(\alpha,\beta; \frac{x}{t}\right)I_{\mathcal{D}}\left(-\frac{x}{t}\right)\biggr\},\label{eq:approximation}
\end{align}
which is demonstrated in Fig.~\ref{fig:2}.
Let $\tilde{\mathcal{D}}$ be the open interval $\left(\sqrt{1-4c^2s^2}\,t,\,\sqrt{1+4c^2s^2}\,t\right)$.
Since we may replace $I_\mathcal{D}(x/t)$ with $I_\mathcal{\tilde{D}}(x)$, the chance of finding the walker in the region $\mathcal{\tilde{D}}$ at time $t$ is extremely small.
And the width of the gap is estimated to become about $2\sqrt{1-4c^2s^2}\,t$ at time $t$.

Due to the limits $\lim_{x\to\sqrt{1+4c^2s^2}} W_{+}(x)=0$ and $\lim_{x\to\sqrt{1-4c^2s^2}} W_{-}(x)=0$, the limit density function generally has four singular points, except for $\theta=\pi/4, 3\pi/4$, that is,
\begin{align}
 &\lim_{x\downarrow \sqrt{1-4c^2s^2}}\,\frac{d}{dx} \lim_{t\to\infty}\mathbb{P}\left(\frac{X_t}{t}\leq x\right)=\left\{\begin{array}{ll}
   +\infty & (\theta\neq\frac{\pi}{4}, \frac{3\pi}{4})\\[1mm]
	  \frac{1}{2\pi} & (\theta=\frac{\pi}{4}, \frac{3\pi}{4})
	 \end{array}\right.,\label{eq:singulality_1}\\
 &\lim_{x\uparrow -\sqrt{1-4c^2s^2}}\,\frac{d}{dx} \lim_{t\to\infty}\mathbb{P}\left(\frac{X_t}{t}\leq x\right)=\left\{\begin{array}{ll}
   +\infty & (\theta\neq\frac{\pi}{4}, \frac{3\pi}{4})\\[1mm]
	  \frac{1}{2\pi} & (\theta=\frac{\pi}{4}, \frac{3\pi}{4})
	 \end{array}\right.,\label{eq:singulality_2}
\end{align}
\begin{align}
 &\lim_{x\uparrow \sqrt{1+4c^2s^2}}\,\frac{d}{dx} \lim_{t\to\infty}\mathbb{P}\left(\frac{X_t}{t}\leq x\right)=+\infty,\label{eq:singulality_3}\\
 &\lim_{x\downarrow -\sqrt{1+4c^2s^2}}\,\frac{d}{dx} \lim_{t\to\infty}\mathbb{P}\left(\frac{X_t}{t}\leq x\right)=+\infty.\label{eq:singulality_4}
\end{align}
Equations~\eqref{eq:singulality_1}--\eqref{eq:singulality_4} are true for any complex numbers $\alpha$ and $\beta$ which satisfy the constraint $|\alpha|^2+|\beta|^2=1$ and determine the initial state of the quantum walk.
Also, we should note that if the value $\pi/4$ or $3\pi/4$ is assigned to the parameter $\theta$, the edges of compact support $\pm\sqrt{1-4c^2s^2}$ take the value $0$ and then the gap around the origin in the probability distribution closes.
These facts are confirmed in Figs.~\ref{fig:4}-(b), \ref{fig:5}, and \ref{fig:2}-(b).

\section{Summary}
Let us summarize this paper.
We took care of a quantum walk whose positions are represented by integer points and analyzed the finding probability as $t\to\infty$.
As a result, one can find that the probability distribution $\mathbb{P}(X_t=x)$ as $t\to\infty$ splits to two major parts, holding a gap around the position where the walker localizes at the initial time.
In the past studies, two quantum walks whose probability distributions could have a gap, were also reported ~\cite{GrunbaumMachida2015,Machida2016b}.
Both papers were studies for limit distributions of time-dependent quantum walks on a line.
Gr{\"u}nbaum and Machida~\cite{GrunbaumMachida2015} analyzed a quantum walk with two inner states and Machida~\cite{Machida2016b} handled a quantum walk with three inner states.
The quantum walk in this study was a different type from the time-dependent quantum walks, but we observed a gap in its probability distribution and the fact was surely demonstrated as the limit law in Theorem~\ref{th:limit}.
Differently from these study, the limit distribution in Theorem~\ref{th:limit} can be determined by either $\alpha$ or $\beta$, which gives the initial state to the quantum walk.
Equations~\eqref{eq:nu+} and \eqref{eq:nu-} actually do not have $\beta$, though their descriptions can be transformed to a description with only $\beta$, removing $\alpha$ with $|\alpha|^2+|\beta|^2=1$.

The study of quantum walks whose distributions hold a gap, is one of the recently interesting topics, and just two kinds of quantum walks with a gap showed up since 2015.
Now another type has been discovered in this paper and we have analyzed its finding probability resulting in the limit distribution provided in Theorem~\ref{th:limit}.

\bigskip
The author is supported by JSPS Grant-in-Aid for Young Scientists (B) (No.16K17648).
\bigskip


\end{document}